\documentclass[11pt]{article}
\usepackage{mons}
\usepackage{graphicx}
\usepackage{times}
\font \sf =cmss10

\newcommand{\teff}{$T_{\mathrm{eff}}$}
\newcommand{\logg}{$\log{g}$}
\newcommand{\metal}{[M/H]}
\newcommand{\etal}{{\it {et al.}\ }}

\lefthead{E. Bertone and A. Buzzoni}
\righthead{Spectral classification of stars using synthetic model atmospheres}

\begin{document}

\title{Spectral classification of stars using synthetic model atmospheres}

\author{Emanuele~Bertone$^{1,2}$ and Alberto~Buzzoni$^{2,3}$\\~\\}
\affil{$^1$Universit\`a degli Studi, via Celoria 16, 20133 Milano, Italy}
\affil{$^2$Osservatorio Astronomico di Brera, via Bianchi 46, 23807 Merate, Italy}
\affil{$^3$Telescopio Nazionale Galileo, A.P. 565, 38700 S/Cruz de La Palma (Tf), Spain}

\begin{abstract}
We devised a straightforward procedure to derive the atmosphere fundamental
parameters of stars across the different MK spectral types by comparing 
mid-resolution spectroscopic observations with theoretical grids of synthetic
spectra.

The results of a preliminary experiment, by matching the Gunn and Stryker 
(1983) and Jacoby \etal (1984) spectrophotometric atlases with the Kurucz
(1995) models, are briefly discussed. For stars in the A-K spectral range,
effective temperature is obtained within a 1-2\% relative uncertainty (at
2-$\sigma$ confidence level). This value raises to 4-5\% for the hottest
stars in the samples (O-B spectral types). A poorer fit is obtained throughout
for stars cooler than 4000~K mainly due to the limiting input physics in the
Kurucz models.
\end{abstract}

\section{Introduction}
A confident estimate of the atmosphere fundamental parameters of stars is of
crucial importance in a number of physical problems dealing with the study of 
the stellar populations in the Milky Way and the other external galaxies.
As a complement of the natural approach, that relies on high-resolution
spectroscopy of single objects, one could also try to take advantage of the
piece of information provided by mid-resolution observations. The latter can
in general probe spectral energy distribution (SED) of stars over a wider
wavelength range, and allow us to collect data in a much shorter observing
time. Physical information on target stars would derive in this case from a
``suitable'' comparison with reference spectra (both theoretical and/or
empirical ones) searching for a best fit over the whole observed SED.

In this work we propose a simple procedure to derive stellar fundamental
parameters (i.e.\ \teff, \logg, \metal) by matching mid-resolution ($\sim
5$~\AA\ FWHM) observations with a grid of theoretical model atmospheres. The
fitting algorithm will be presented in Sec.~2 while some preliminary results
of our method will be discussed in Sec.~3.

The Gunn \& Stryker (\cite{gun83}, hereafter GS83) and the Jacoby \etal
(\cite{jac84}, JHC84) spectrophotometric atlases provided the wide sample of
target stars (336 in total) for our experiment, with different spectral type
(from O to M) and MK luminosity class (from I to V). This allowed us to test
our procedure across the whole HR diagram.

The 175 GS83 SEDs are defined in 503 points between 3160 and 10620 \AA, with a
step of 10 or 20~\AA. This coarse sampling basically sets also the intrinsic
spectral resolution of the data. Absolute fluxes are given in the AB magnitude
scale (Oke \& Gunn 1983). The 161 JHC84 SEDs are sampled in 2799 points
between 3510 and 7427 \AA, by steps of 1.4 \AA. They have been obtained at a
resolution of 4.5 \AA\ FWHM. Both samples of stars were originally corrected
for interstellar reddening.

As a reference framework for our procedure, we relied on the ATLAS~9 synthetic
models by Kurucz (\cite{kur95}).  Other theoretical databases could in
principle be used, of course, like for instance the recent and more elaborated
NextGen 5 models by Hauschildt \etal (\cite{hau99a}, \cite{hau99b}; see
Bertone \etal 2001 for a critical discussion in this regard).

The adopted ATLAS~9 grid consists of 409 models in total, assuming a solar
metallicity, a microturbulent velocity of 2 km/s and a
mixing-length-to-scale-height ratio L/H$_p$=1.25.  Effective temperature spans
the range $50\,000 \geq T_\mathrm{eff} \geq 3500$~K with a step in the model
grid of 250~K for stars cooler than 10\,000~K, increasing up to 2500~K for
warmer stars. Surface gravity explores the range $0.0 \leq \log{g} \leq
5.0$~dex at steps of 0.5~dex. All these model atmospheres and spectra are
available on the Internet at the Kurucz www site ({\sf
http://cfaku5.harvard.edu/}).

\section{The fitting algorithm}

Our method basically consists in a measure of the likelihood function, $s$,
that quantifies the similarity between target spectrum and each template SED
of the reference grid. As far as the quantity $s$ is known across the grid,
that is as a function of the physical parameters of the model atmospheres, a
formally ``best'' solution is identified by the minimum of $s$ in the (\teff,
\logg, \metal) parameter space.  The underlying hypothesis of this choice is
of course that similar spectra are produced by similar physical parameters in
a univocal way.

Operationally, the fiducial atmosphere parameters for the {\it i}-th star in
our target sample are obtained in three steps:

{\it i)} we first compute a residual function, $X$ vs. $\lambda$ in the
flux logarithm domain such as
\begin{equation}
X_{(i,j)}(\lambda) = [\ln f_i(\lambda) - \ln f_j(\lambda)]\ W(\lambda)
\end{equation}
where $f_i$ is the SED of the target star, $f_j$ is that of the {\it j}-th
reference model atmosphere and $W$ is a weighting factor that only depends on
$\lambda$. In case a wavelength resampling is necessary to consistently match
target and template spectra, this will always be done by rebinning the
template spectrum at the wavelength values of the target. We also assume that
the template spectrum has been preliminarily degraded to the same resolution
of the target data.

{\it ii)} We therefore compute the standard deviation of the statistical
variable $X$ within the wavelength range of the target observations:
\begin{equation}
\sigma(X)_{(i,j)} = \sqrt{{\it Var(X)}}
\end{equation}
As in eq.~(1) we worked in the natural logarithm domain, $\sigma(X)_{(i,j)}$
can be read, to a first approximation,  as a mean percent deviation of the
{\it j}-th template spectrum with respect to the {\it i}-th target spectrum.

{\it iii)} For each template we therefore compute the corresponding likelihood 
function, $s$, defined as
\begin{equation}
s_{(i,j)} = \sigma(X)_{(i,j)}.
\end{equation}
The minimum of the quantity $s_{(i,j)}$ across the grid of reference spectra will
eventually identify the ``best'' fiducial atmosphere parameters for the {\it
i}-th target star. As a further refinement, to increase the accuracy of our
fitting algorithm, we actually search for a minimum of the $s_{(i,j)}$
function after a spline smoothing.

\begin{figure*}
\begin{center}
\begin{tabular}{cc}
\includegraphics[width=7.5cm]{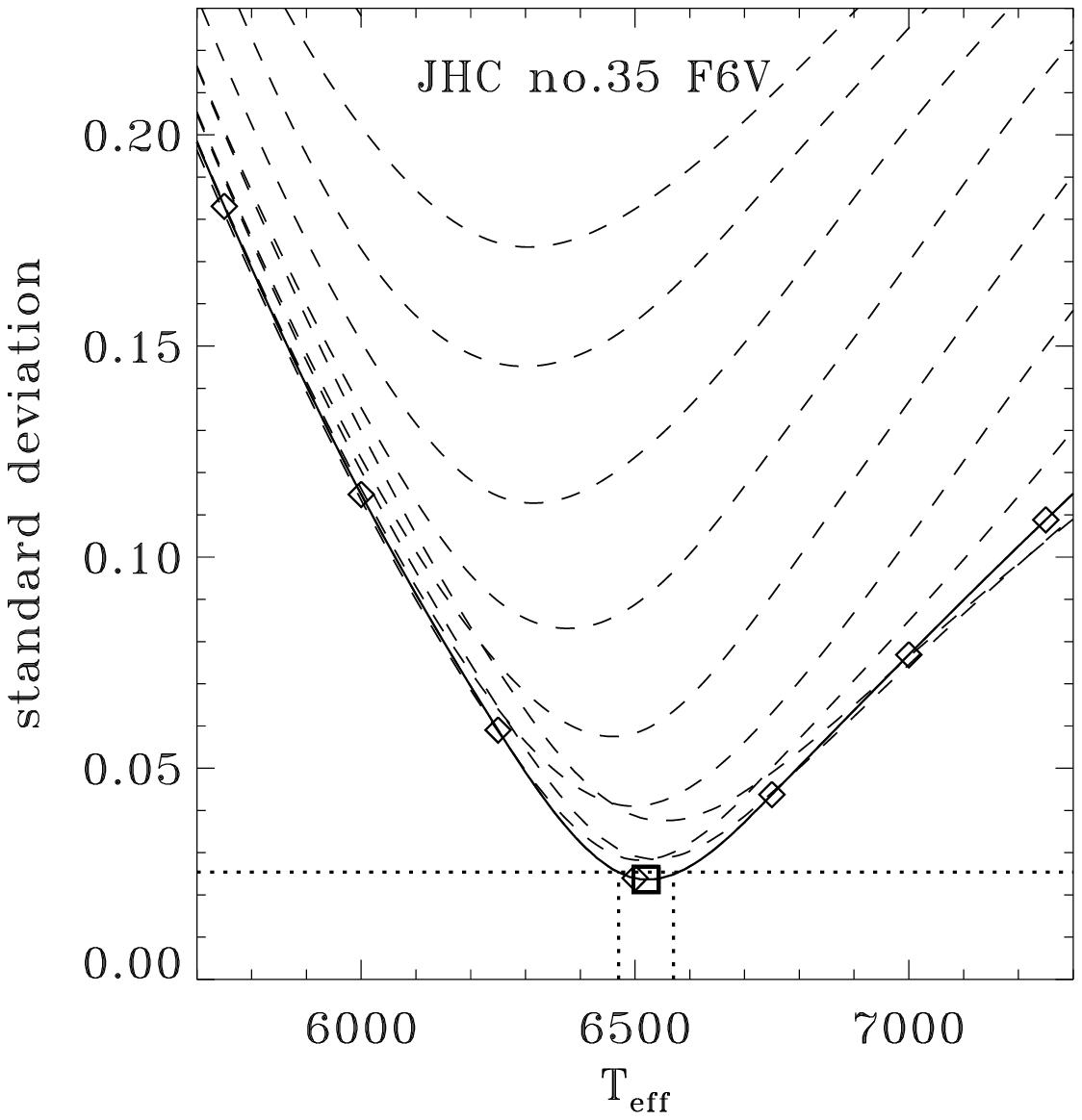} &
\includegraphics[width=7.5cm]{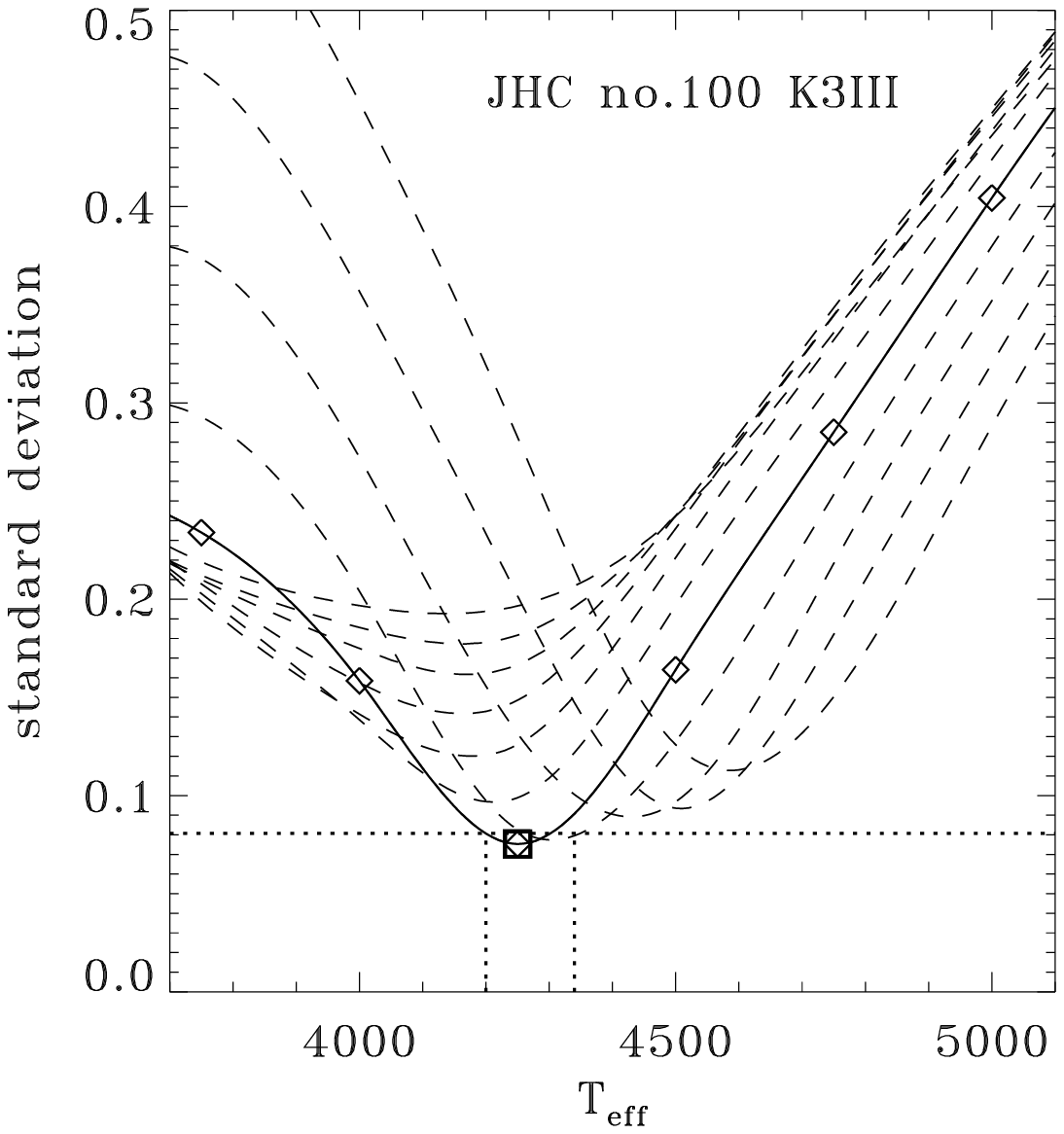} \\
\end{tabular}
\end{center}
\vspace{-4mm}
\caption[]{\label{fig:spline35-100} {\small Spline interpolation of the loci
of the points $s$ vs. \teff\ for two stars in the JHC84 atlas. The solid line
is the curve at the {\it best} fiducial gravity ($\log g = 4.0$ for JHC84
no.35 and $\log g = 2.0$ for JHC84 no.100); the dashed lines are the
curves at the other surface gravity values; diamonds are the interpolated $s$
values at the {\it best} gravity (the points at other gravities have the same
abscissas); the square indicates the minimum of the curve. The horizontal
dotted line shows the upper limit $s_\mathrm{max}$ of the 2-$\sigma$
confidence interval of the standard deviation, given by the $F$ test. This
limit defines the \teff~error range, indicated by the vertical dotted
lines. The error on \logg\ is provided by the gravity value of the curves that
pass under $s_\mathrm{max}$.}}
\end{figure*}

An example of our procedure for star no.\ 35 (an F6V dwarf) and no.\ 100 (a
K3III giant) in the JHC84 atlas is shown in Fig.~\ref{fig:spline35-100}.
In order to estimate the statistical uncertainty in the fiducial set of
atmosphere parameters, for each target star we performed an $F$ test on the
likelihood function $s$ (the freedom degrees in this case are the number of
SED points accounted in the fit). The resulting 2-$\sigma$ confidence interval
for a one-tail $F$ test then translates into a confidence range for \teff~and
\logg, as displayed in Fig.~\ref{fig:spline35-100}.

\begin{figure*}
\begin{center}
\begin{tabular}{cc}
\includegraphics[width=7cm]{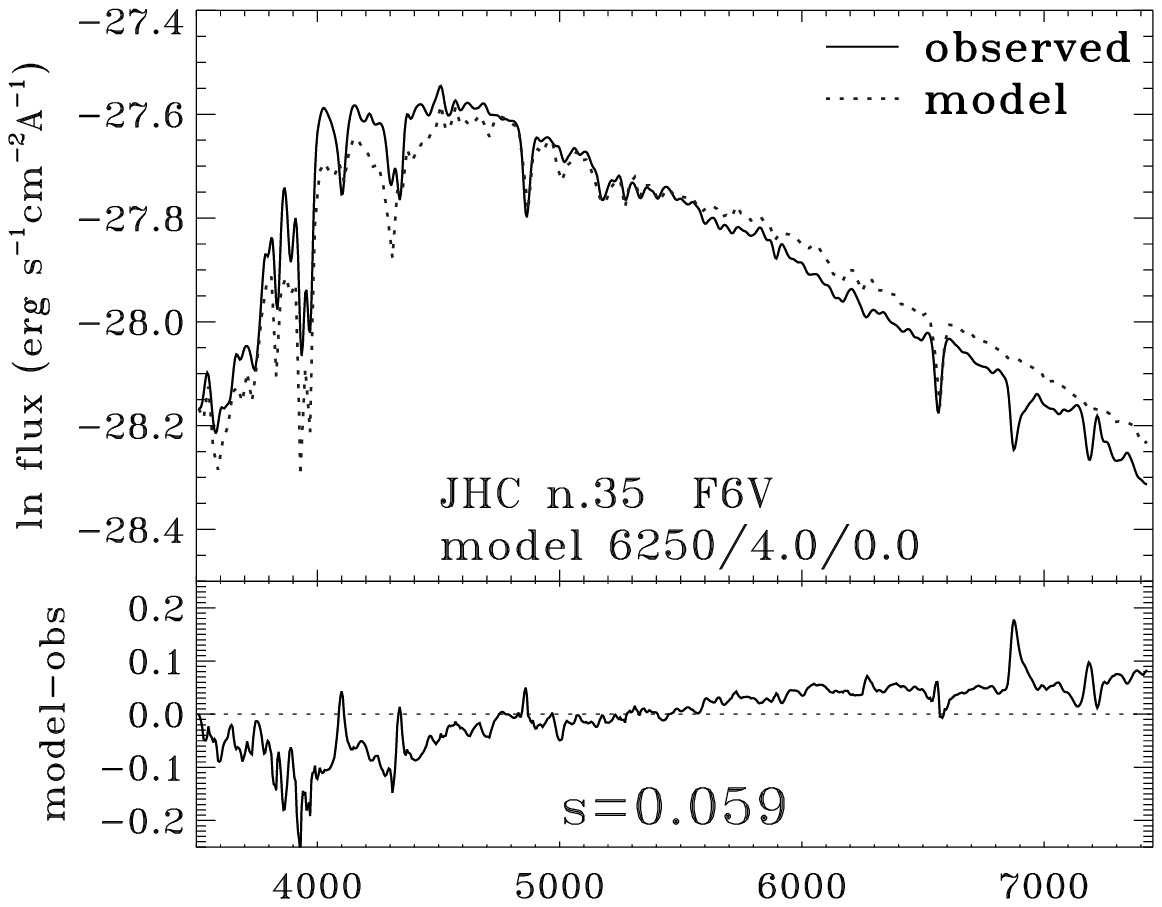} &
\includegraphics[width=7cm]{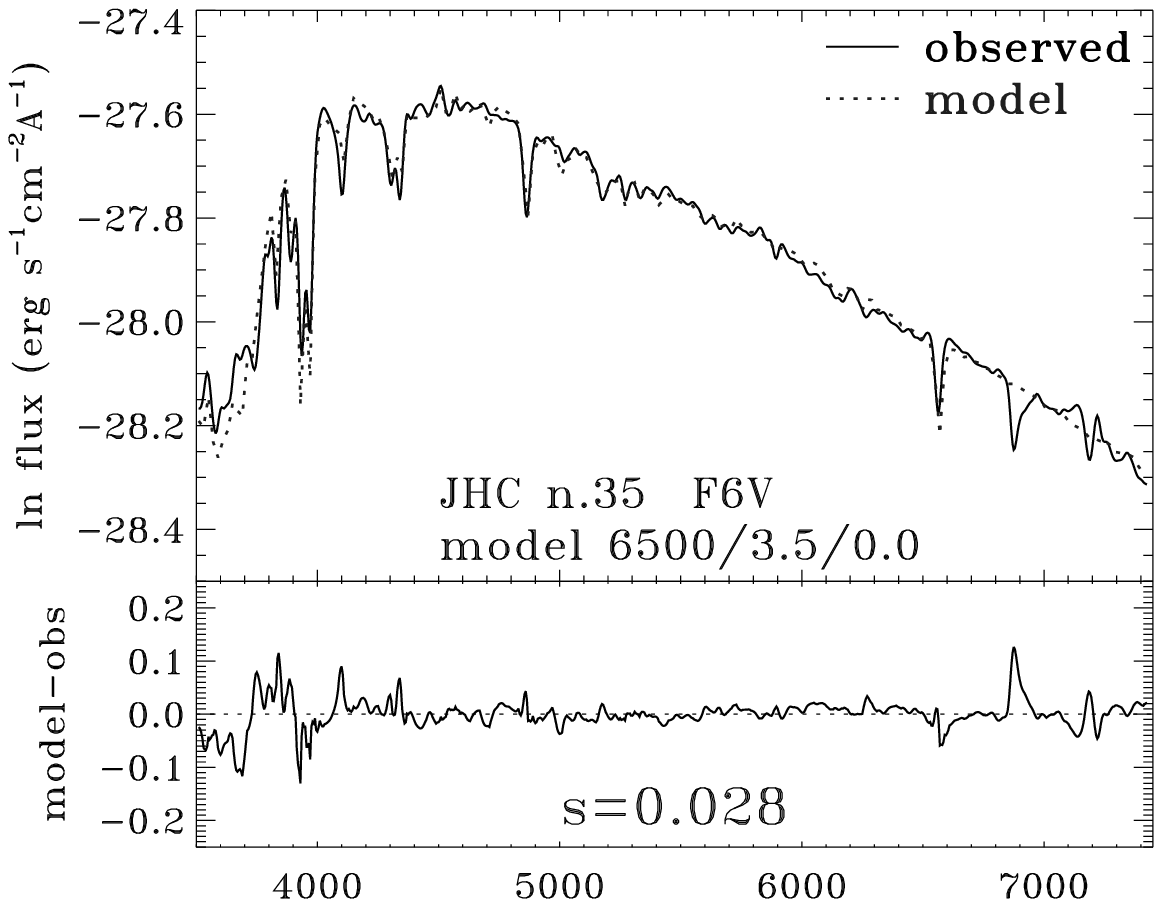} \\
\multicolumn{2}{c}{\includegraphics[width=7cm]{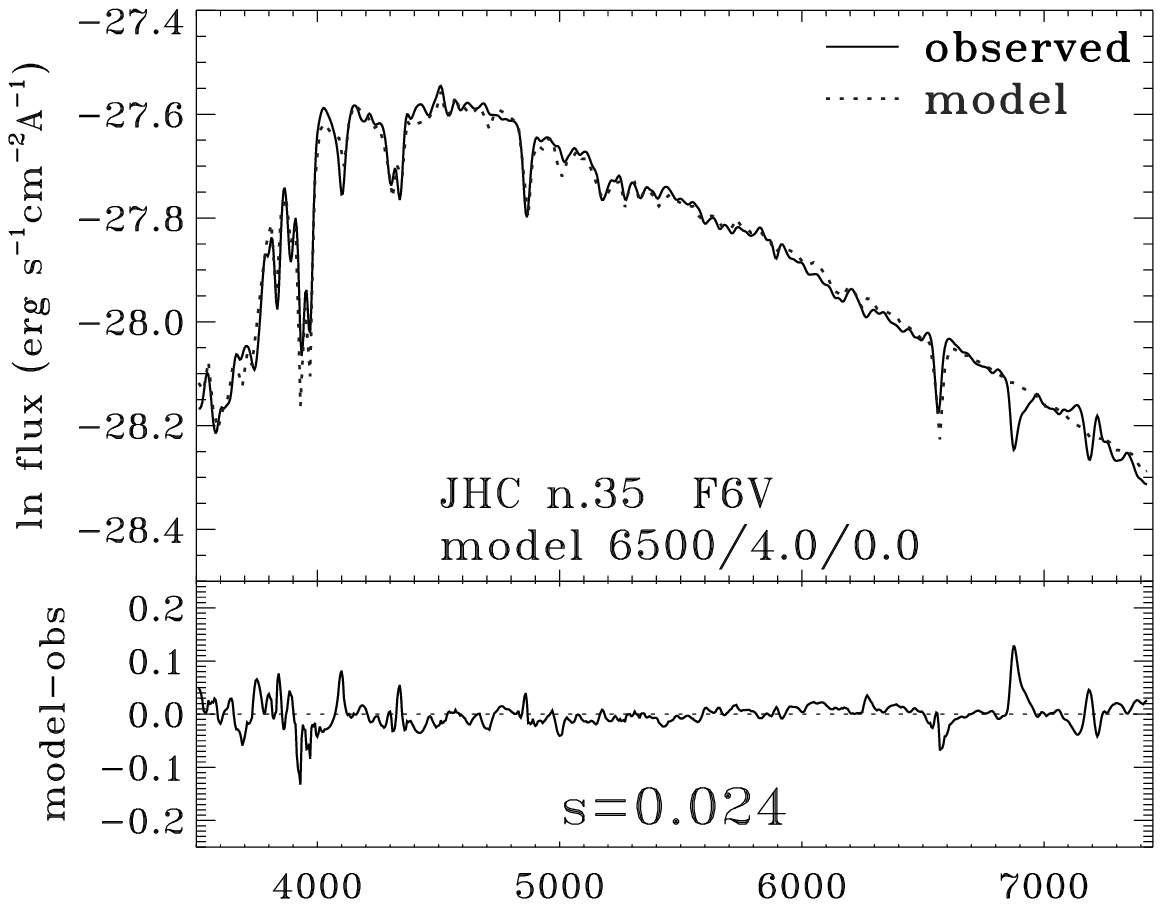}} \\ 
\includegraphics[width=7cm]{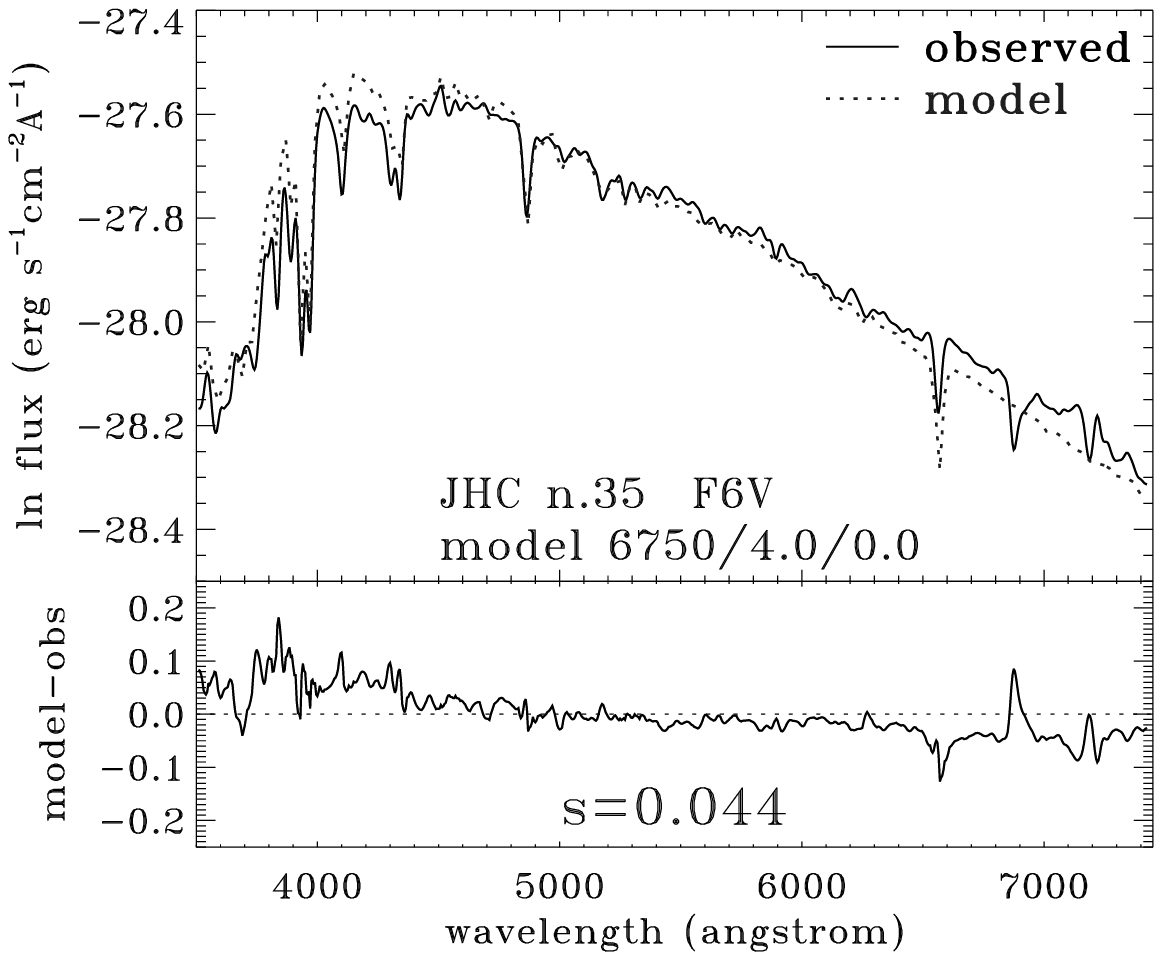} &
\includegraphics[width=7cm]{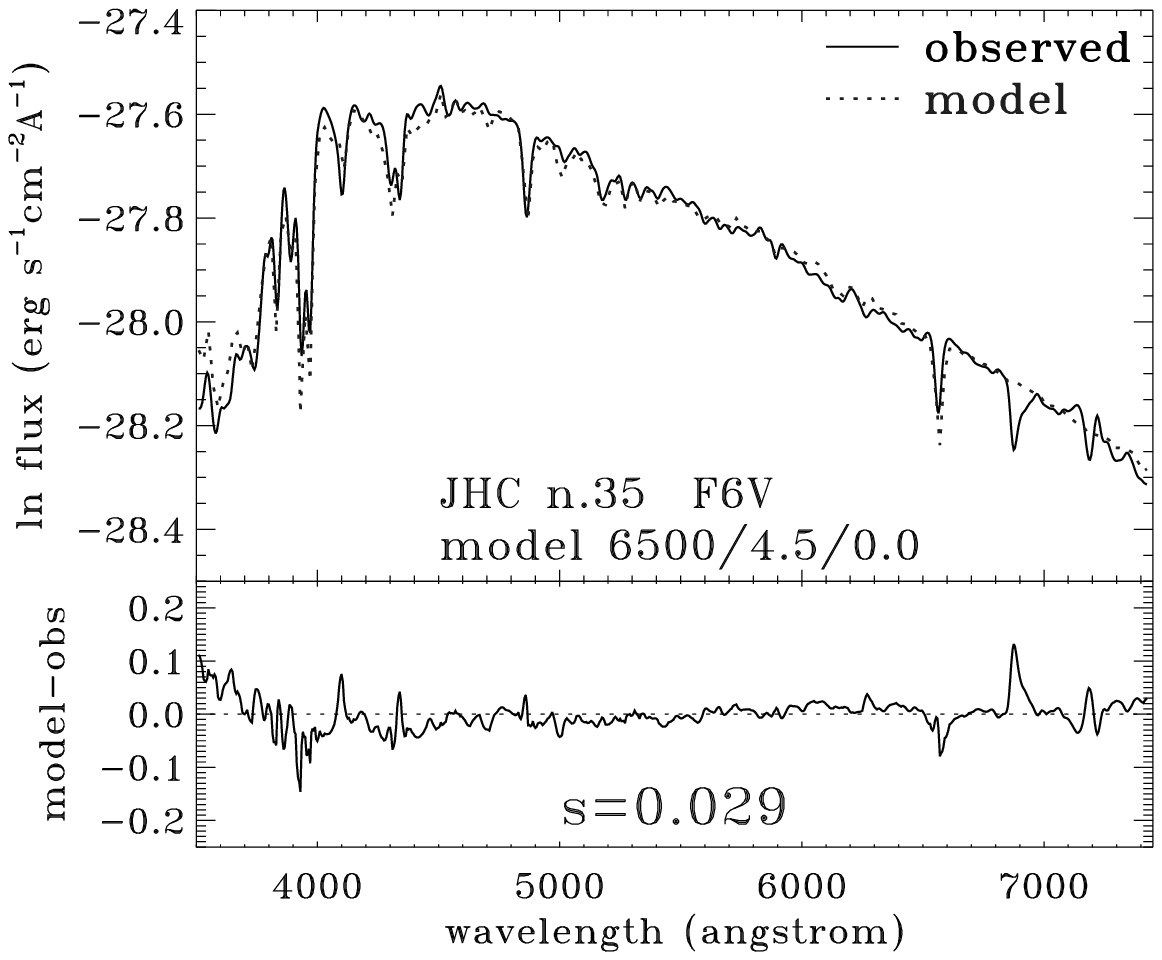} \\
\end{tabular}
\end{center}
\vspace{-4mm}
\caption[]{\label{fig:comparison} {\small The SED of the star JHC84 no.35 (SAO
57199, F6$\,$V) is compared with several ATLAS9 SEDs. In the left column, the
models have same gravity and $\Delta$\teff=250 K. In the right column, the
models have same \teff~and $\Delta$\logg=0.5 dex. The lower panels show the
value of $X_{(i,j)}(\lambda)$ (see eq.(1)).}}
\end{figure*}

Full details of the matching procedure for the same star no.\ 35 of
Fig.~\ref{fig:spline35-100} are also reported in Fig.~\ref{fig:comparison}, by
comparing with Kurucz templates with different temperature (left panels) and
surface gravity (right panels). The central panel in the figure is the best
fit for \teff = 6500~K and \logg = 4.0 dex, assuming a solar metallicity.

As far as the continuum emission is concerned, note that the trend in the
residual flux is much more sensitive to a change in temperature while any
change in gravity would especially reflect in the ultraviolet emission
shortward of the Balmer break.

Some artificial features appear in the GS83 spectra about 7600 \AA\ and 9400
\AA. They are due to telluric absorption and have therefore been excluded in
our analysis by setting W=0 in eq. (1) for the relevant wavelength. In
addition, also the region shortward of 3500 \AA\ has been rejected for the
same sample due to a poorer flux calibration of the original data in the
ultraviolet wavelength range.

\section{Results and discussion}

As our method entirely relies on the Kurucz synthetic models, its performance
is mainly determined by their properties, virtues and faults. The ATLAS~9
models are LTE, plane-parallel and they use the mixing-length theory with
overshooting for the treatment of convection. A huge amount of atomic and
molecular lines are considered, but triatomic molecules are not implemented
in the opacity calculations, and Kurucz (1993) himself suggests not to use
these models for M stars.

For 293 out of 336 stars in the GS83 and JHC84 atlases our procedure converged
to an acceptable set of fiducial atmosphere parameters (namely \teff\ and
\logg, assuming a solar metallicity). About 85\% of these stars have
$s_\mathrm{min} < 0.07$, with a mean of 0.05 for the total sample.
All stars with poorer fit (i.e. $s_\mathrm{min} > 0.11$) are cooler than
4000~K. For 43 stars no solution was found as $s_\mathrm{min}$ presumably
located beyond the physical boundaries of the model grid. They are mostly O
and M stars, that is at the extreme edges of the temperature scale.
Molecular contribution seems the most likely responsible for the trouble with
M stars, while for O stars this seems rather to deal with the non-LTE regime
affecting stellar atmospheres at hotter temperature.

\begin{figure*}[ht]
\begin{center}
\includegraphics[width=9cm]{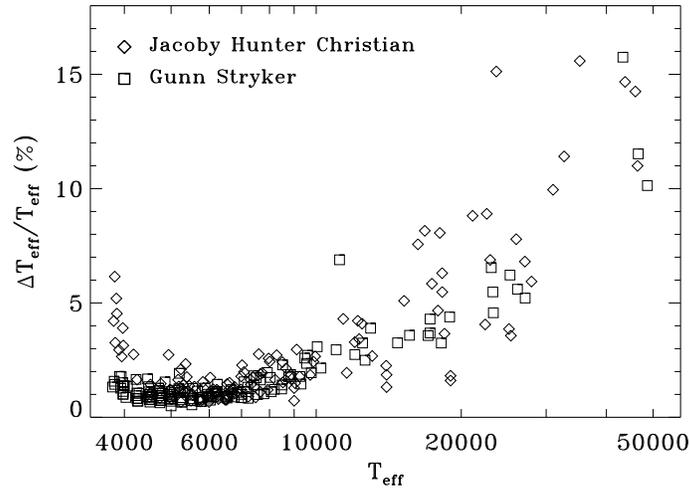}
\end{center}
\vspace{-4mm}
\caption[]{\label{fig:tefferror} {\small Relative error on the \teff\ of the
stars of the two atlases. The typical 2-$\sigma$ absolute error is
$\pm$50--100~K for K-F stars, $\pm$100--200~K for A stars and $\pm$400--1000~K
for B stars.}}
\end{figure*}

Figure~\ref{fig:tefferror} gives a summary of our results by reporting the
relative uncertainty in the derived \teff\ for the 293 stars with fitting
solution. The GS83 stars provided slightly better results compared with the
JHC84 sample because of a wider wavelength baseline of the spectra that
allowed a more accurate determination of  \teff.
The tipical relative uncertainty (at 2-$\sigma$ confidence level) in the
temperature estimate for A-K stars turns about $\pm$~1-2\%, raising to 4-5\%
for B stars. The typical error box for the surface-gravity estimates is $\pm
0.5$ dex.

As a final remark, it is worth stressing that our procedure could in principle
work also for high-resolution spectra. Different problems would come out in
this case, however, deserving to be preliminarily assessed in order to assure
a confident fit to the data.

 From the operational point of view we know, in fact, that the calibration of 
{\it echelle} spectra at a high-order dispersion is a more critical task to
perform over a wide wavelength baseline compared to corresponding
low-resolution observations.
Computation of theoretical spectra at high-resolution is on the other hand
very time consuming, and template grids are usually not available for public
release. Furthermore, they take a huge amount of disk space and often cannot
be so easily managed with the current computing facilities.

A suitable compromise in this sense could rely on a selective analysis of
narrow-band spectrophotometric indices, like in the the Lick system (Worthey
\etal \cite{wor94}) or according to Rose's (\cite{ros94}) studies, taking full
advantage in this case of the high-resolution piece of information but
restraining the analysis only to much reduced (and strategically placed)
windows to probe the stellar SED.


\begin{thebibliography}{}

\bibitem[2001]{ber01}
	Bertone E., Buzzoni A., Chavez M., \& Rodriguez L., 2001, in Proc. of ``The link between stars and cosmology'', eds. M. Chavez, A. Bressan, A. Buzzoni and D. Mayya, Kluwer, Dordrecht, in press
\bibitem[1983]{gun83}
	Gunn J.E. \& Stryker L.L., 1983, ApJS, 52, 121
\bibitem[1999a]{hau99a}
	Hauschildt P.H., Allard F. \& Baron E., 1999, ApJ, 512, 377
\bibitem[1999b]{hau99b}
	Hauschildt P.H., Allard F., Ferguson J., Baron E. \& Alexander D.R., 1999, ApJ, 525, 871
\bibitem[1984]{jac84}
	Jacoby G.H., Hunter D.A. \& Christian C.A., 1984, ApJS, 56, 257
\bibitem[1993]{kur93}
	Kurucz R.L., 1993, in Proc. IAU Symposium 149, eds. B. Barbuy and A. Renzini, Kluwer, Dordrecht, p.225
\bibitem[1995]{kur95}
	Kurucz R.L., 1995, CD-ROM n.13, revised
\bibitem[1983]{oke83}
	Oke J.B. \& Gunn J.E., 1983, ApJ, 266, 713
\bibitem[1994]{ros94}
	Rose J.A., 1994, AJ, 107, 206
\bibitem[1994]{wor94}
	Worthey G., Faber S.M., Gonz\'alez J.J. \& Burnstein D., 1994, ApJS, 94, 687
\end{thebibliography}
\end{document}